# Interlacing in atomic resolution scanning transmission electron microscopy

Jonathan J. P. Peters[1, 2], Tiarnan Mullarkey[1, 2, 3], James A. Gott[4, 5], Elizabeth Nelson[2], Lewys Jones[1, 2, 3]

[1] Advanced Microscopy Laboratory, Centre for Research on Adaptive Nanostructures & Nanodevices (CRANN), Trinity College Dublin, Dublin, Ireland

[2] School of Physics, Trinity College Dublin, Dublin, Ireland

[3] Centre for Doctoral Training in the Advanced Characterisation of Materials, AMBER Centre, Trinity College Dublin, Dublin, Ireland

[4] Department of Physics, University of Warwick, Coventry, UK

[5] Advanced Materials Manufacturing Centre (AMMR), Warwick Manufacturing Group (WMG), University of Warwick, Coventry, UK

## Abstract

Fast frame-rates are desirable in scanning transmission electron microscopy for a number of reasons: controlling electron beam dose, capturing in-situ events or reducing the appearance of scan distortions. Whilst several strategies exist for increasing frame-rates, many impact image quality or require investment in advanced scan hardware. Here we present an interlaced imaging approach to achieve minimal loss of image quality with faster frame-rates that can be implemented on many existing scan controllers. We further demonstrate that our interlacing approach provides the best possible strain precision for a given electron dose compared with other contemporary approaches.

## Introduction

Scanning transmission electron microscopy (STEM) has become a widespread and powerful technique in materials science, particularly with the development of practical aberration correctors. An electron beam is focussed to a spot, with a diameter as small as $< 100$ pm (Sasaki et al., 2012), and is scanned across the surface of a material. As the beam is scanned, or rastered, across the material, multiple signals can be collected simultaneously using different detectors. For example, images can be formed based on scattering to annular dark field (ADF) or spectroscopic information can be retrieved from electron energy loss spectroscopy (EELS) or energy dispersive X-ray (EDX) spectrometers (Muller, 2009). Increasingly, segmented or pixelated detectors are being used to obtain more elaborate contrast mechanisms (Ophus, 2019).

The ability to gather these multiple signals at atomic resolution is one of the main merits of STEM. However, the serial fashion of the scanning process is relatively slow compared to parallel HRTEM. In conventional scanning, this drawback is compounded by an additional flyback wait time at the beginning of each scan line, used to allow for lens hysteresis effects to decay. There is then another similar wait time between each frame, though this may also compensate for limited data streaming bandwidth or other delays. The total time for an image to be acquired, $T_{Total}$, can then be summarised as

$$T_{Total} = \delta_t n_x n_y + T_{LFB} n_y + T_{FFB}, \tag{1}$$

where $\delta_t$ is the pixel dwell-time, $n_x$ and $n_y$ are the number of pixels in x and y respectively, $T_{LFB}$ is the line flyback time, and $T_{FFB}$ is the frame flyback time (Mullarkey et al., 2022). For example, a relatively fast but typical scan consisting of 512×512 pixels, 1 μs pixel time and 500 μs line flyback takes over 0.5 s per image, yielding a maximum frame-rate of 1.9 frames per second (fps).

Limited scan speeds cause a number of problems. Firstly, non-rigid distortions from environmental distortions or specimen drift contort crystallography in an image and reduces both resolution and strain mapping fidelity. Conveniently, the conventional rectangular scanning used in STEM allows for practical diagnosis and removal of non-rigid distortions from sequentially acquired images (Jones et al., 2015; Berkels et al., 2014). Another effect of the slow scan speed is the associated beam damage due to the highly concentrated electron beam(Egerton, 2019). Typical strategies to circumvent this are to acquire multiple images at a faster rate, with the same total dose incident on the specimen, but with the dose rate significantly reduced (Jones et al., 2017). However, with faster scanning, the signal to noise of individual images is reduced and the reduction in dwell time is limited by the scan and detection hardware (Ishikawa et al., 2020; Mullarkey et al., 2021; Mittelberger et al., 2018). Similarly, the beam current can simply be reduced, though again giving an adverse effect on signal to noise of individual images (Buban et al., 2010). Whilst image series can be reconstructed and averaged to gain signal, this does not help capture dynamic events in the microscope, for example in in-situ experiments, where the poor frame-rate necessarily limits the temporal sensitivity of experiments.

Recent work has shown that the line flyback time can be significantly reduced, without detriment to the retrievable image quality, from typical values of 500 μs to as low as 20 μs (Mullarkey et al., 2022). The remaining option to increase imaging frame-rates further is to reduce the number of pixels in an image. Typically, this simply results in a reduced resolution or smaller field of view. A summary of current approaches to STEM acquisition and the corresponding efficiency (i.e. the percentage of acquisition time that produces image data) and fps is given in Table 1. Around ten years ago, the field of compressed sensing (CS) emerged as an approach to reduce beam damage and decrease frame times. CS uses novel scan paths or high-speed beam blanking to only acquire a subset of pixels in an image, later reconstructing the remainder using computer algorithms (Stevens et al., 2014). There is some doubt as to the advantages of this method (especially when low-dose data becomes Poisson limited) (Van den Broek et al., 2019; Sanders & Dwyer, 2020), though one definite limitation is the requirement for an advanced scan generator or beam blanker (Béché et al., 2016; Li et al., 2018). Uncertainty also surrounds the ability to use CS for quantitative position analysis, as the intrinsically sparse nature of the data means scan-distortion errors are not readily diagnosed for use with non-rigid registration.

For inspiration to improve scan speeds in current STEM systems, we turn to the early 20$^{th}$ century (Kell et al., 1936). Interlaced video became a widespread technology where only half the lines in a video frame are displayed at once (Figure **1**). Usually frames alternate between showing even and odd lines where, either by human perception or computational means, the missing lines can be reconstructed to recover a full frame from each interlaced frame as shown in Figure **1**b. The most important aspect of this was, that for the same frame-rate and resolution, interlaced video required only half the bandwidth, or storage capacity, as compared to a fully sampled video. Alternatively, for any given transmission bandwidth or storage capacity a doubling of frame-rate can be achieved. This has not gone unnoticed in the electron microscopy community, with interlaced scanning used in scanning electron microscopy (SEM) to limit charging effects (Postek, 1984). However, the differences in resolution and implementation can make this tricky to simply transfer to high-resolution STEM imaging. Most notably, images should be acquired close to Nyquist frequency for an efficient use of beam dose, interlacing may result in acquiring images below Nyquist frequency and reconstructing the missing lines from interlacing (i.e. deinterlacing) may not be straightforward. In STEM it is also not often possible to alternate between capturing even and odd lines, meaning that video deinterlacing algorithms are not always applicable. On the other hand, interlacing affords us the opportunity to reduce the electron dose by at least a factor of two and, unlike compressed sensing, is compatible with existing scan generators and non-rigid registration methods.

Here we present the application of interlaced scanning and digital super-resolution methods to atomic resolution STEM imaging. We quantitatively evaluate and compare multiple deinterlacing algorithms in the context of single frame images, image stack averaging and image video series. With these results, we present a robust and practical strategy that is applicable to many STEM instruments in use today.

## Methods

### Experimental interlacing

Interlacing is often not readily available on most STEM systems and controllers. If the STEM scan controller has functionality for a custom scan, for example the point electronic DISS interface, the interlaced scan can be directly programmed. However, scan controllers that do not directly support such functionality can be made to perform an interlaced scan with relative ease, an advantage of interlaced scans vs other more complicated advanced scans. The majority of scan controllers, e.g. Gatan's Digiscan II, have some functionality to control the scan gains (amplifications), usually used to finely tune the scan magnification in x and y. By halving the gain of the non-interlaced scan direction and doubling the number of pixels in the same direction, an interlaced scan can be formed. This approach is trivial to implement but does have some limitations in not being able to realise scan rotation through the scan controller, as is useful for non-rigid registration approaches (Sang & LeBeau, 2014). It is also somewhat non-ideal to need to change advanced technical settings between users in a multi-user facility, and could lead to miscalibration of the magnification if set incorrectly.

### Deinterlacing

One of the advantages of modern software and programming languages, such as python, is the abundance of publicly available code and libraries for a range of image processing, for example OpenCV (Bradski, 2000), SciPy (Virtanen et al., 2020), and DIPlib (Luengo & Et al, 2021). The range of potential deinterlacing algorithms span a range of complexities, from simple line double to interpolation and advanced in-painting algorithms. Each method has a unique set of advantages and disadvantages, where some may be more robust to noise, others may provide a better reconstruction of images samples close to Nyquist frequency. The implementation of each method used here has been made available in a GitHub for the reader to use or adapt.

### Strain Precision

Interlacing provides an interesting opportunity for non-rigid registration to achieve higher strain precisions for a given electron dose. Jones et al. (Jones et al., 2017) demonstrated that, for a fixed total dose, the strain precision is proportional to $1/\sqrt{n}$, where $n$ is the number of frames. In theory, interlacing doubles the number of frames for a given dose and should outperform an equivalent dose of full frame imaging by a factor of $1/\sqrt{2}$. To test this hypothesis, we have developed an approach to incorporate realistic environmental image distortions into simulated image. This allows us to apply the same *time dependent* distortions (a more realistic approach than pixel dependent) to a series of full frame or interlaced images. To approximate the distortion field, a number of sine waves (typically

50-100) are chosen with random amplitudes, frequencies, phases, and directions. These are then summed and applied to an image in the *time domain* (e.g. accounting for pixel dwell time, flyback wait time, and frame number in the series) using spline interpolation to resample the image. Strains can then be measured using the geometric phase analysis (GPA) method (Hÿtch et al., 1998) within the Strain++ software (Peters, 2021). An example of a perfect image of $SrTiO_3$ and its synthetically distorted version are shown in Figure **2**a, with the corresponding error in strain demonstrated in Figure **2**b. Both the full frame and interlaced simulated and distorted image series can then be non-rigidly aligned and the strain precision measured and fairly compared. It should be noted that there is also potential for interlacing to help with images limited by a low signal to noise ratio from low dose. Instead of doubling the number of frames, the dwell time of each pixel could be doubled to provide an improvement in signal to noise for fixed total frame time and beam dose.

## Results and Discussion

### Deinterlacing Comparison

A deinterlacing method suitable for STEM data must first be chosen. To test the various deinterlacing algorithms, we first take a full frame image and synthetically interlace it, deinterlace it and compare the result to the original full frame ground truth. The different between the deinterlaced image and the ground truth reveals any limitations of the deinterlacing approach, as shown in Figure **3**a. At the same time, we can measure the algorithm's speed with the goal of determining if real-time deinterlacing is possible under typical experimental conditions.

We tested and compared a range of widely available algorithms, with a subset of popular algorithms shown in Figure **3**b. These include: (1) Line doubling, where each line is simply repeated twice; (2) Bilinear interpolation, assuming interpolated points lie on a linear gradient between adjacent points; (3) Custom bilinear interpolation, where information is incorporated from the fast scan direction; (4) Lanczos interpolation, a resampling method that can preserve sharp edges (Duchon, 1979); (5) Bicubic interpolation, using polynomials to interpolate data between nearest points. (6) FFT boxcar resampling, removing frequencies in the Fourier transform above the interlacing frequency; (7) Telea inpainting provided by OpenCV (Telea, 2004); (8) Navier Stokes inpainting provided by OpenCV (Bertalmio et al., 2005). The output of a more extensive set of algorithms is shown and compared in supplementary Figure S1. The code to reproduce Figure **3** has been made available on GitHub (https://github.com/TCD-Ultramicroscopy/STEM-deinterlacing), so that the reader can examine the deinterlacing algorithms as well as test with their own algorithms, their own images, or their own hardware.

Whilst the exact results of Figure **3** can vary between images, and we encourage the reader to test the algorithms for their specific imaging conditions and needs, some preferred approaches start to emerge. For example, line doubling stands out as the fastest method, with bicubic interpolation giving the best error from the ground truth. There are then compromises such as bilinear interpolation, with speeds closer to line doubling, but with an error closer to bicubic. Because of this, and the ease of implementation in different programming languages, bilinear interpolation has been chosen as the default deinterlacing method for the work presented here. Perhaps disappointingly, the advanced inpainting algorithms give some of the highest error for the lowest speeds, though it should be noted that the computation speed is somewhat subjective as it depends on the required scanning speed as well as the specific hardware. Equally the speed depends on the specific algorithm, for example, the two bicubic interpolation implementations from the DIPLib (Luengo & Et al, 2021) and SciPy (Virtanen et al., 2020) libraries have drastically different speeds. Further to this, large improvements in speed might be realised using multithreading or graphical processing units.

### Frame-rates

The maximum possible frame-rate improvement from interlacing is a factor of two. However, frame flyback can limit the achievable framerate, potentially consisting of data streaming from the scan controller to the computer, copying/saving data, and factors internal to the scan generator. To compare the experimental frame-rates improvements from interlacing, a summary of frame times is given in Table 2. We have tested an FEI Titan G2, equipped with a point electronic DISS scan control, as well as a Nion UltraSTEM 200 and JEOL ARM200F, both equipped with a Gatan Digiscan II. The frame-rates from the Digiscan II only show a speed improvement of %64 when interlacing. This is due to an extra inter frame time from Eq. 1 measured to be 0.154 ± 0.005 s, independent of scan size, that limits the achievable frame-rate. The point electronic DISS controller significantly reduced the inter frame time, allowing interlacing to achieve a 99.5% increase in frame-rate compared to full frame imaging. In any case, interlacing gives a significant improvement in imaging frame-rate for minimal penalty.

### Strain precision

To compare the strain precision of images with varying dose fractionation but equal total dose, an artificial, strain free image of [100] SrTiO$_3$ was generated as a series of gaussians at 512×512 pixel size (a detailed multi-slice simulation is not necessary for this analysis). Pixel dwell times, $\delta_t$, and total number of frames, $n$, were chosen for full frames and interlaced frames such that $\delta_t \cdot n$ was 40 μs and 80 μs, respectively. Distortions representing the effects of environmental disturbances were then generated (with behaviours matched fairly with respect to time rather than scan-location) and applied both to the full frame and interlaced series with and without 90° scan rotation between frames.

Interlaced images were then deinterlaced using bilinear interpolation and each image series was then non-rigidly aligned with SmartAlign (Jones et al., 2015), using the same settings throughout. The strain error, which on a perfect single crystal is any deviation from 0% strain, was then quantified as the standard deviation of the $\varepsilon_{yy}$ strain component (Hÿtch et al., 1998), where the $\varepsilon_{yy}$ component was chosen as most of the strain error is expected in the slow y-scan direction compared to the fast x-scan direction, assuming no scan rotation.

If we consider that the apparent probe-displacements resulting from randomised uncorrelated scan distortions have a Gaussian-like distribution (Jones et al., 2017), then following frame averaging the width of the distribution describing these would be expected to reduce with a $1/\sqrt{n}$ relationship (where n is the number of frames averaged). The resulting strain error of the full frame and interlaced series are shown in Figure **4**a for simulated data as a function of the number of dose equivalent full frames (i.e. the number of experimentally recorded probe positions divided by the number of image pixels). The $1/\sqrt{n}$ proportionality is immediately visible from the fit lines in all cases.

The imaging strategy with the largest strain error is the full frame imaging with no scan rotation between frames, as would be expected from the literature. The equivalent interlaced strain precision is a factor of 1.59 ± 0.07 lower (better), exceeding the expected $\sqrt{2}$ (1.41) improvement. This better than expected improvement may be a result of halving the apparent strain gradient in the deinterlaced image due to the two times faster sampling of the distortion field in the interlaced approach. The reduced apparent strain gradient is likely better diagnosed and corrected by the non-rigid registration in SmartAlign. Interestingly the full frame images with scan rotation also give a similar 1.61 ± 0.09 improvement compared to full frame imaging, perhaps because of the orthogonal redundant views of the same sample. A further 1.26 ± 0.03 improvement in the strain precision is found by combining scan rotation with the interlaced acquisition, giving a total improvement of 2.0 ± 0.1 from the full frame series without rotation. The strain precision improvement from interlaced to interlaced with rotation of 1.264 ± 0.005 suggest that, whilst still a significant improvement, interlacing benefits less from scan rotation. This is possibly because the $\varepsilon_{yy}$ strain component is already improved from partially measuring the fast scan direction and interlacing only provides an improvement to the frames where the slow scan direction is collinear with the measured strain orientation. In any case, the maximum strain precision for a given dose budget is achieved when using STEM interlacing with scan rotation between frames.

As the simulated measurements of Figure **4**a are under synthetic conditions with purely Poisson noise behaviour, these findings were verified experimentally. Figure **4**b shows the equivalent measurement of strain precision from a JEOL ARM200F controlled by a Digiscan II. As mentioned earlier, the Gatan Digiscan II hardware configuration does not support automated acquisition of interlaced images with scan rotation. Similar to previous works by Jones et al. (Jones et al., 2017), at higher dose fractionation (the most noisy images) the strain precision does not continue to decrease endlessly with the expected $1/\sqrt{n}$ proportionality. A discussion of this effect has been presented in the supplementary materials, with a probable cause being higher frequency environmental distortions. In this case, further improvements to strain precision necessitate improved microscope or room/environment designs (Muller et al., 2006). Despite this limitation, an improvement in strain precision is still achieved, with a factor of 1.4 ± 0.1 improvement again agreeing with the expected value of $\sqrt{2}$. With our experimental measurements, it can be seen that a dose fractionation using interlacing of 20 frames (equivalent to 10 full frames) provides the best strain precision for a fixed dose. The reader is encouraged to reproduce this optimisation for their own conditions (for example sample material or detector angles) but this serves an a reasonable starting condition.

## Conclusion

We have demonstrated the applicability and utility of interlacing to atomic resolution STEM imaging using simple, readily available, and computationally efficient algorithms. Acquiring interlaced STEM images does not require further financial investment in hardware or software, providing an accessible and sustainable method to increasing the capabilities of any existing STEM. Interlacing provides an easy way to double imaging frame-rates, reduce dose-rate, and allow dynamic in-situ events to be captured whilst remaining compatible with existing non-rigid registration techniques. The methods explored here are easily implemented and can also provide benefits to spectroscopic or 4D imaging. In particular, interlaced STEM combined with scan rotation and non-rigid registration achieved the highest strain precision for a fixed total dose budget compared to conventional full frame imaging.

### Code and Data Availability

The code which is used to test deinterlacing methods, add distortions to images, and measure strain can be found at our Github repository https://github.com/TCD-Ultramicroscopy/STEM-deinterlacing. The data used in this paper can be found at doi: 10.5281/zenodo.7137495.

## Acknowledgements

The authors would like to acknowledge the Centre for Research on Adaptive Nanostructures and Nanodevices (CRANN) and the Advanced Materials and BioEngineering Research (AMBER) Network for financial and infrastructural support for this work. In particular we acknowledge the help and support of the Advanced Microscopy Laboratory (AML) staff and facilities. J.J.P.P. and L.J. acknowledge

SFI grant 19/FFP/6813, L.J. acknowledges SFI/Royal Society Fellowship grant URF/RI/191637, T.M. acknowledges the SFI & EPSRC Centre for Doctoral Training in the Advanced Characterisation of Materials (award references 18/EPSRC-CDT-3581 and EP/S023259/1). J. A. G acknowledges EPSRC grant EP/P031544/1. L.N. as an undergraduate at the time acknowledges the support of the Trinity College Dublin School of Physics.

## Conflict of Interest

The authors declare no competing interests.

CONTRIBUTORS (2020). SciPy 1.0: fundamental algorithms for scientific computing in Python. *Nature methods* **17**, 261–272.

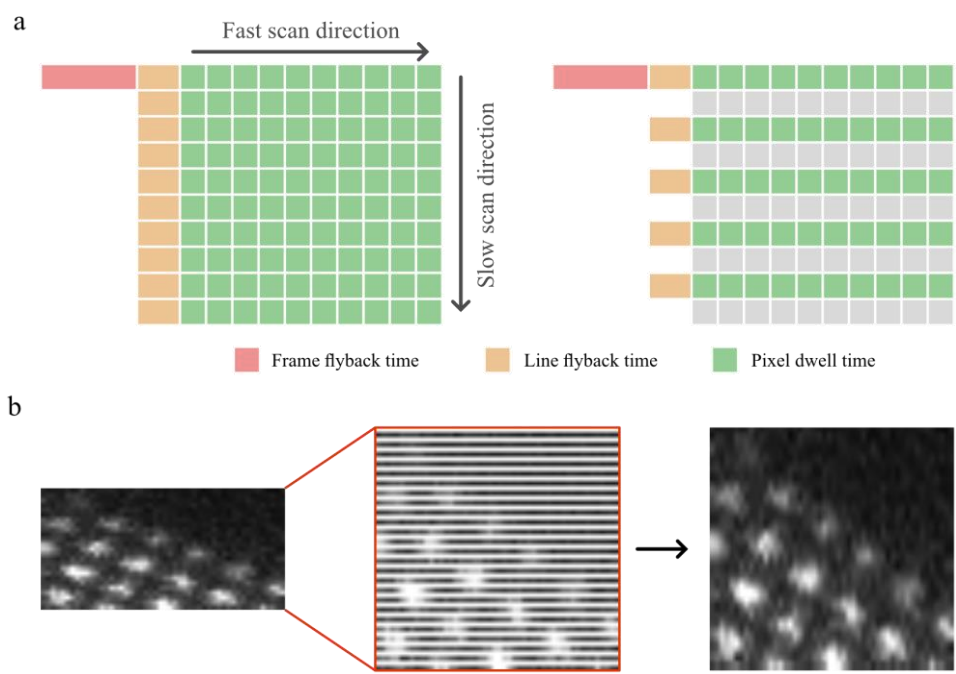

**Figure 1 a** Schematic of full frame and interlaced acquisitions showing the breakdown of total frame time. **b** Raw experimental interlaced ADF image of Au columns. **c** shows **b** with the scan lines spaced correctly in real space and **d** shows the image after deinterlacing using bilinear interpolation.

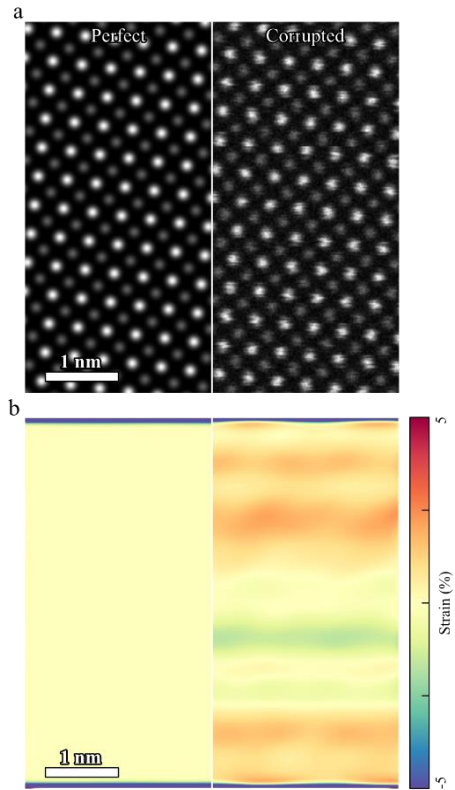

**Figure 2 a** Example simulated image shown before and after distortions have been introduced (scan-corrupted). **b** $e_{yy}$ Strain measurement from the perfect and corrupted image shown in **a**.

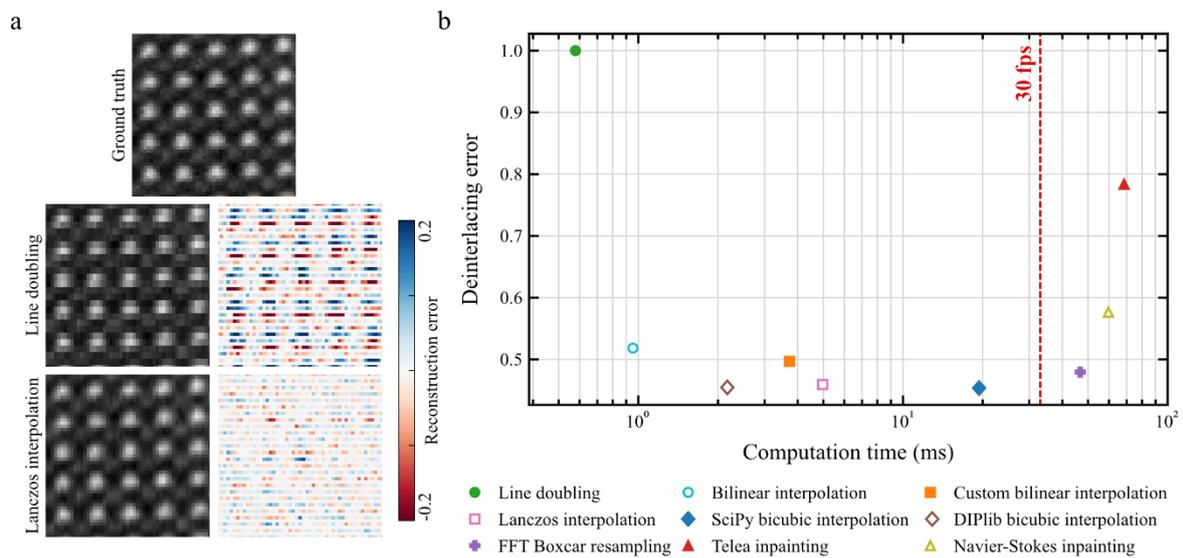

**Figure 3 a** Ground truth experimental ADF image of PbZr$_{0.2}$Ti$_{0.8}$O$_3$ with exemplar line-doubled and Lanczos deinterlaced images. The difference to the ground truth is shown as a fraction of the intensity range of the ground truth image. **b** Deinterlacing error versus computation time. Deinterlacing error is calculated as the root mean square error. Times are an average of 1000 calculations performed on an Intel i7-10700. The ground truth image in **a** is a crop of the image used to calculate the deinterlacing error in **b**. The full image and expanded plot of **b** can be found in the supplementary materials.

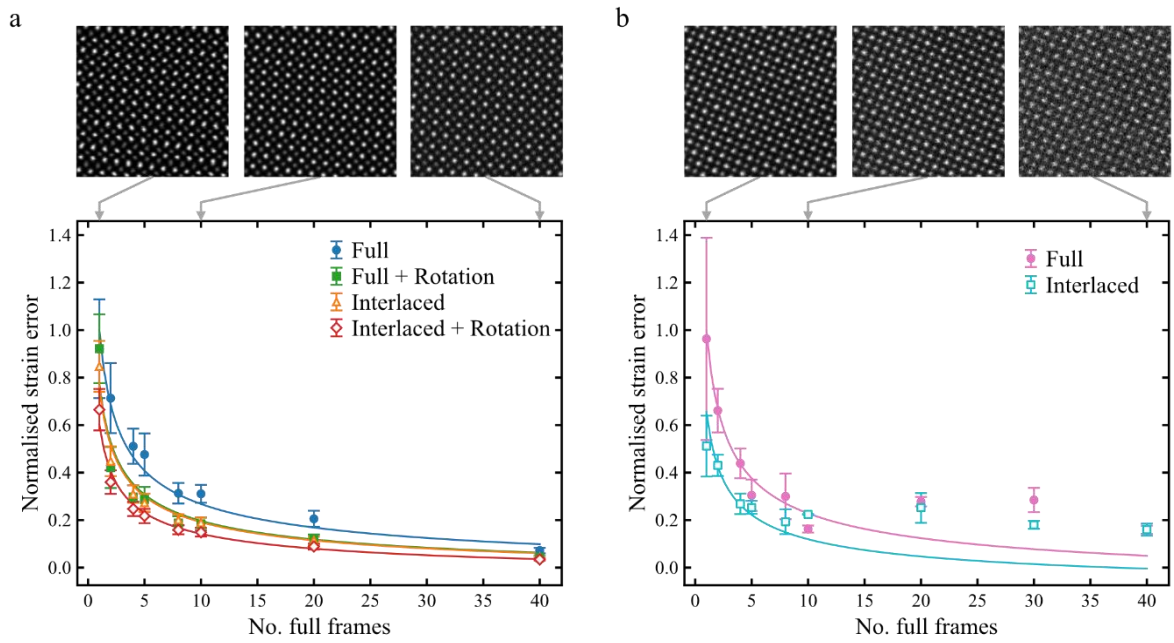

**Figure 4** Strain error of reconstructed images as a function of dose fractionation and across different scanning strategies. **a** and **b** show simulated and experimental data, respectively, with example single frames show above. Fit lines are $\propto 1/\sqrt{n}$ and error bars show the standard error of the means.

**Table 1** Overview of scanning acquisition strategies with corresponding acquisition efficiency and achievable frames per second. The acquisition efficiency is defined as the fields of view captured per frame time dose exposure, with frame time accounting for beam exposure not contributing to data acquisition (e.g. line flyback time).

| Acquisition Strategy | Motivation | Max. Acquisition Efficiency (%) | Max. Achievable fps |
|---|---|---|---|
| Conventional line-sync (50 Hz) and flyback (500 µs) at 512x512 px | Ease of use, line-sync mitigates A.C. mains image distortions | 97.49 | 0.098 |
| Reduced dwell-time (1 µs) followed by non-rigid registration | Faster scanning reduces frame time and effects of environmental distortion | 50.59 | 1.93 |
| Reduced dwell-time and reduced flyback time (20 µs) | Reduced flyback waiting improves beam-efficiency | 96.24 | 3.67 |
| Interlacing (this work) | | 192.48 | 7.34 |
| Reduced pixel dimensions (256x256 px) | Reduced number of pixels further reduces frame-time | 92.75 | 14.15 |
| Interlacing and reduce image dimensions | | 185.51 | 28.31 |

**Table 2** Expected and experimentally achieved frame-rates for conventional and interlaced imaging with a given set of scan parameters.

| Pixel dwell time (µs) | Line flyback time (µs) | Scan width (pixels) | Scan Height (pixels) | Scan controller | Expected fps | Achieved fps |
|---|---|---|---|---|---|---|
| 3 | 500 | 300 | 300 | Gatan Digiscan II | 2.38 | 1.74 ± 0.02 |
| | | | 150 | | 4.72 | 2.85 ± 0.03 |
| | | | 300 | point electronic DISS 6 | 2.38 | 2.3701 ± 0.0006 |
| | | | 150 | | 4.76 | 4.728 ± 0.004 |

# Supplementary materials for: Interlacing in atomic resolution scanning transmission electron microscopy

Jonathan J. P. Peters[1, 2], Tiarnan Mullarkey[1, 2, 3], James A. Gott[4], Elizabeth Nelson[2], Lewys Jones[1, 2, 3]

[1] Advanced Microscopy Laboratory, Centre for Research on Adaptive Nanostructures & Nanodevices (CRANN), Trinity College Dublin, Dublin, Ireland
[2] School of Physics, Trinity College Dublin, Dublin, Ireland
[3] Centre for Doctoral Training in the Advanced Characterisation of Materials, AMBER Centre, Trinity College Dublin, Dublin, Ireland
[4] Department of Physics, University of Warwick, Coventry, UK

## Deinterlacing Comparison

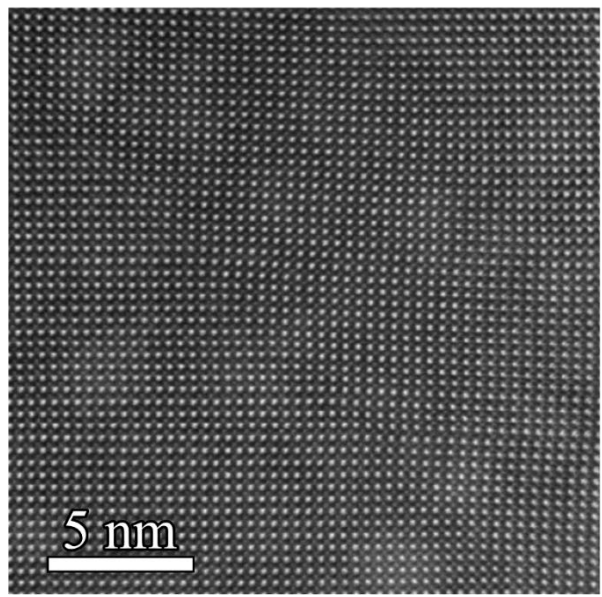

**Figure S1** Raw image of $PbZr_{0.2}Ti_{0.8}O_3$ used for comparing deinterlacing methods in Fig. 3

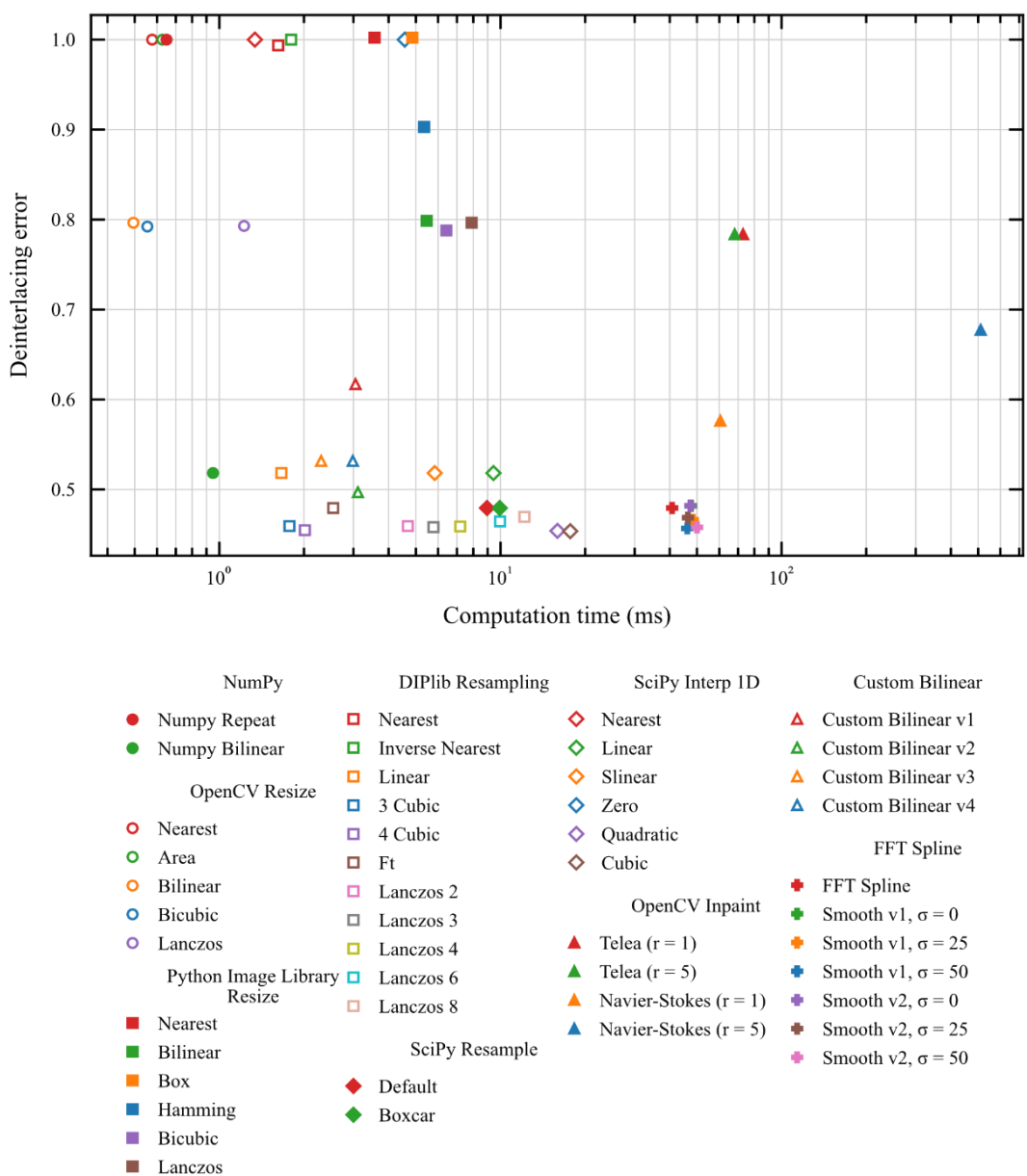

**Figure S2** Comparison of deinterlacing methods using Fig. S1 as a ground truth starting image. Times are an average of 1000 calculations performed on an Intel i7-10700.

For testing the interlacing methods, the image in Fig. S1 was used as a representative atomic resolution with sampling close to Nyquist frequency. Figure S2 shows a more extensive version of Fig. 3 in the main text. The algorithms consist of custom written approaches and those from external libraries which are shown in full in the GitHub repository (https://github.com/TCD-Ultramicroscopy/STEM-deinterlacing)

## Strain precision

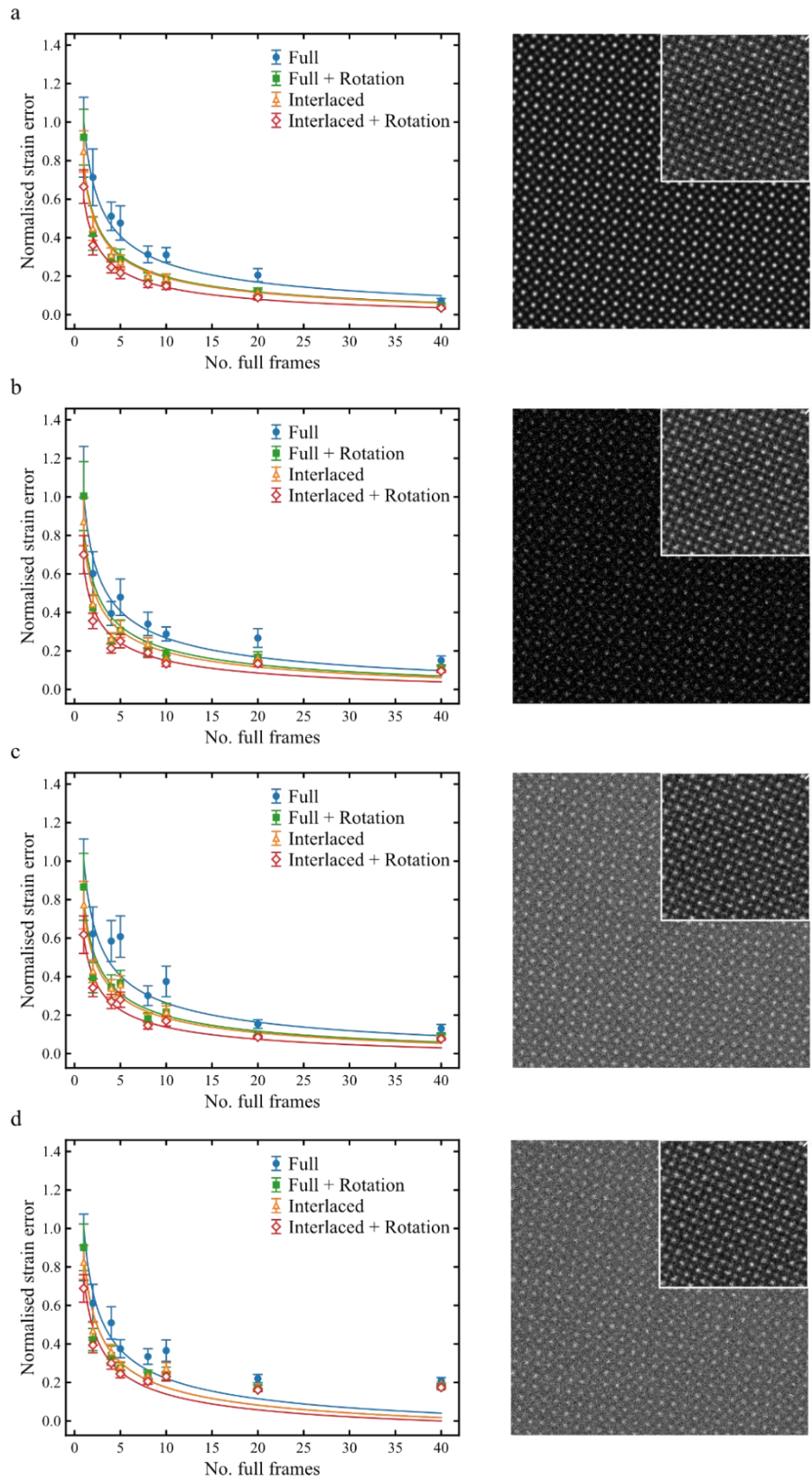

**Figure S3** Strain error measurements for high signal to noise, **a**, Poisson noise, **b**, Gaussian noise, **c**, and amplified high frequency noise, **d**. The accompanying images show an example frame from the 40 frames series data. Inset is the corresponding experimental image.

The experimental strain precision measurements shown within the main text in Fig. 4b (reproduced in Fig. S3a for convenience) and within the work of Jones et al. (Jones et al., 2017) show a deviation from the expected strain precision as a function of dose fractionation (i.e. splitting the total dose between a number of frames to be registered and averaged.). A $1/\sqrt{n}$ proportionality is predicted, but at higher frame counts the strain error is higher than expected. Using our newly developed simulation approach to measuring the strain error, we explored possible causes of the aforementioned strain error behaviour.

One theory proposed by Jones et al. (Jones et al., 2017) was the visibility of streaking from slow scintillator streaking. However, we do not observe this at most of the scan speeds used here, with only the fastest scanning speeds (1 μs per pixel) exhibiting a small amount of streaking. For this reason we discount this reasoning.

The other explanation presented is the lack signal to noise ratio, particularly from read-out noise (i.e. Gaussian noise). Figure S3b and Fig. S3c show the simulated strain precisions with the inclusion of additional Poisson and Gaussian noise, respectively. In both cases there is no significant decrease in the strain precision for higher frame numbers, despite the signal to noise ratio being worse than the equivalent experimental image (as shown in the inset images). However, if the amplitude of higher frequency noise is increased, the experimental behaviour is better reproduced, as shown in Fig. S3d. This suggests that environmental factors limit the strain precision, and the microscope or room design should be improved to further improve strain precision.